\documentclass[%
preprint,
 amsmath,amssymb,
 aps,
]{revtex4-2}

\usepackage{graphicx}
\usepackage{dcolumn}
\usepackage{bm}
\usepackage{physics}
\usepackage{soul}

\begin{document}


\title{\textit{Ab initio} description of $\bar{p}+\rm{^3H}$ and $\bar{p}+\rm{^3He}$ systems in optical models}
\author{Pierre-Yves Duerinck}
\email{pierre-yves.duerinck@ulb.be}
\affiliation{Physique Nucléaire Théorique et Physique Mathématique, C.P. 229, Université libre de Bruxelles (ULB), B-1050 Brussels, Belgium. \\ IPHC, CNRS/IN2P3, Université de Strasbourg, 67037 Strasbourg, France.}

\author{Rimantas Lazauskas}
\email{rimantas.lazauskas@iphc.cnrs.fr}
\affiliation{IPHC, CNRS/IN2P3, Université de Strasbourg, 67037 Strasbourg, France.}


\date{\today}

\begin{abstract}
In the context of the ongoing PUMA experiment (CERN), which investigates antiproton annihilation on atomic nuclei, we study the energy shifts and widths of Rydberg states in the $\bar{p}+\rm{^3H}$ and $\bar{p}+\rm{^3He}$ systems by performing \textit{ab initio} calculations. The scattering lengths and scattering volumes are first determined by solving the Faddeev–Yakubovsky equations in configuration space. The level shifts and widths of the corresponding $\bar{p} \, \rm{^3H}$ and $\bar{p} \, \rm{^3He}$ hydrogen-like states are then obtained using the Trueman formula. A pronounced model dependence associated with the nucleon–antinucleon interaction is observed for certain states. Finally, annihilation densities are computed from the four-body wavefunctions. Comparison with the nuclear density distributions indicates that the nucleon–antinucleon annihilation is predominantly peripheral.


\end{abstract}

\maketitle


\section{Introduction}
CERN is the only facility worldwide to provide low-energy antiproton beams using the Antiproton Decelerator (AD) and the Extremely Low ENergy Antiproton ring (ELENA). This unique capability allows the high-precision spectroscopy of antiprotonic atoms \cite{ALPHA,ASACUSA,GBAR}, which are formed when an antiproton is captured by an atom or ion. The capture typically occurs into a highly excited Coulomb orbital with principal quantum number $n \approx 40$. The further cascade of the antiproton to lower energy states occurs via x-ray emission and is accompanied by the successive depletion of the electronic shells via Auger electron emission.  Ultimately, when the antiproton reaches low-lying states, it annihilates with a nucleon via the strong interaction, producing mesons in a predominantly pion-rich final state.

 This mechanism will be exploited in the PUMA experiment \cite{PUMA_A22}, which aims to probe the nuclear density tails of stable and radioactive isotopes. Low-energy antiprotons have previously been shown to provide a sensitive probe into the outer regions of nuclear densities \cite{EH_40yearofpbar,TJ01,KT07}, and thus provide information on the nuclear surface. In the PUMA experiment, measurements of the annihilation products following the decay of antiprotonic atoms are therefore expected to provide reliable information on the tails of proton and neutron density distributions.

From a theoretical perspective, a microscopic description of antiproton–nucleus systems is challenging, as it requires detailed knowledge of both nucleon–nucleon ($NN$) and nucleon-antinucleon ($N \bar{N}$) interactions. While $NN$ interaction models are well constrained by abundant data, the $N \bar{N}$ interaction remains poorly known. The highly complex annihilation dynamics is usually modelled using optical potentials~\cite{BP68,DR80,LL82,KW86,ELLW09,Julich}. However, the limited experimental data available to constrain these models makes it difficult to assess their reliability. A solid understanding of hydrogen-like antiproton–nucleus states is essential for interpreting PUMA’s results, and it is important to test the stability of theoretical predictions against uncertainties in the $N \bar{N}$ interactions. 

Some of the first theoretical studies related to the PUMA project focused on assessing the model dependence of physical observables when using existing optical potentials in two- and three-body systems. A comparison of $N \bar{N}$ phase shifts performed in Ref. \cite{CHW23} revealed significant discrepancies among the considered models. Furthermore, investigations of the antiproton–deuteron ($\bar{p}d$) system in Refs. \cite{CL21,*DLC23,DLD23} showed a strong model dependence in the energy levels of  $P$ states, while $S$ states were found to be comparatively model-independent. In the present paper, we report the first four-body \textit{ab initio} calculations for the $\bar{p}+\rm{^3H}$ and $\bar{p}+\rm{^3He}$ systems employing realistic $NN$ and $N \bar{N}$ potentials.

The attractive Coulomb interaction between the nucleus and the antiproton gives rise to an infinite set of hydrogen-like Rydberg states. For a nucleus with mass number $A$ and charge $Z$, the energy of these hydrogenic states is given by
\begin{equation}
E^{(C)}_{n} = B_A + \epsilon_n, \quad \epsilon_n = -\frac{Z^2}{n^2} R_y(\bar{p}A),
\end{equation}
where $B_A$ is the binding energy of the nucleus. The corresponding Rydberg constant $R_y(\bar{p}A)$ and Bohr radius $B(\bar{p}A)$ are 
\begin{equation}
R_y(\bar{p} A) = \frac{1}{2} \mu_{\bar{p}A} c^2, \quad B(A \bar{p})= \frac{\hbar c}{Z \alpha \mu_{A \bar{p}}},
\end{equation}
where $\mu_{\bar{p}A}$ is the reduced mass of the $\bar{p}A$ system and $\alpha$ is the fine structure constant.


In addition to relativistic and QED effects, which are not considered in this work, the strong interaction induces both a shift and a broadening of the hydrogenic levels. A key observable is the level shift, defined as
\begin{equation}
\Delta E_{nl} = E_{nl} - E^{(C)}_n = \Delta E_r - i \frac{\Gamma}{2},
\end{equation}
where $E_{nl}$ denotes the energy of the state under consideration. One of our goals is to compute the level shifts of the low-lying $\bar{p} \, \rm{^3H}$ and $\bar{p} \, \rm{^3He}$ states, where annihilation is most likely to occur. To this end, we solve the four-body Faddeev-Yakubovsky equations \cite{Y67,LC20}  in configuration space. The zero-energy scattering problem is solved for different states, and the corresponding level shifts are extracted using Trueman’s series expansion~\cite{T61}, truncated at second order:
\begin{align}
\frac{\Delta E_{nl}}{\epsilon_n} = - 4 \, \frac{ \alpha_{nl}}{n} \frac{a_l}{B^{2l+1}} \left(1- \beta_{nl} \frac{a_l}{B^{2l+1}} \right),  \label{Tr}
\end{align}
where $a_l$ is the calculated scattering length and where the coefficients $\alpha_{nl}$ and $\beta_{nl}$ are given in Ref. \cite{CRW92} for $l=0,1$. This approach has previously proven effective for the $p \bar{p}$ and $\bar{p}d$ states \cite{CRW92,CL21,*DLC23,DLC25}.

The paper is organized as follows. First, the framework of $N \bar{N}$ optical models and the formalism of Faddeev-Yakubovsky equations are presented. The level shifts of $S$ and $P$ states in the $\bar{p} \, \rm{^3H}$ and $\bar{p} \, \rm{^3He}$ systems are then computed using different $NN$ and $N \bar{N}$ interactions to investigate the model dependence. The results are compared with previous theoretical predictions and available experimental data. Finally, the annihilation region is investigated by computing the annihilation densities associated with low-lying Rydberg states.

\section{Formalism}
\subsection{The $N \bar{N}$ interaction}

The scarcity of experimental data on the $N \bar{N}$ systems motivates to exploit similarities with the $NN$ interaction. 
Within meson-exchange theory, the elastic component of the $N \bar{N}$ interaction can be related with $NN$ potentials through a $G$-parity transformation \cite{K02}. However, some meson-exchange terms -- particularly those involving one pion exchange -- become singularly attractive after the G-parity transformation, requiring additional regularization to obtain physically meaningful potentials.

The G-parity transformation does not account for the $N \bar{N}$ annihilation. At rest, the large amount of available energy leads to annihilation into a vast number of many-body meson-producing channels, which prevents a rigorous microscopic treatment. As a result, the $N \bar{N}$ annihilation is typically described using simplified phenomenological models. The most common approaches are coupled-channel \cite{SHP85_CCNNbar,LF90_CCNNbar,YC21} and optical potential \cite{K02} frameworks. We demonstrated that these two approaches yield very similar predictions for the $ \bar{N}d$ system at low energy, provided that they are phase-equivalent in the $N \bar{N}$ sector. For this reason, we adopt the simpler optical-model framework in the present study.

Traditional optical potentials employ phenomenological Woods–Saxon form with parameters fitted to experimental data, as in the Bryan–Phillips \cite{BP68}, Dover-Richard \cite{DR80}, and Kohno-Weise \cite{KW86} models. An alternative strategy has been used by the J\"ulich group \cite{Julich}, whose potential incorporates G-parity–transformed pion-exchange contributions derived from chiral effective field theory, while the annihilation is represented by complex contact terms. The model parameters are adjusted to reproduce the $N\bar{N}$ phase shifts of the Nijmegen partial-wave analysis \cite{T12_NPWA} or directly fit existing $N\bar{N}$ scattering data. 

For $\bar{p}A$ systems, the long-range interaction is dominated by the Coulomb force. Yet, $N \bar{N}$ potentials are usually formulated in the isospin basis, which is not well suited for the inclusion of electromagnetic effects. Furthermore, nucleon mass differences lead to distinct asymptotic behaviours in
$n\bar{p}$, $p\bar{n}$, $n\bar{n}$ and $p\bar{p}$ channels. It is therefore more appropriate to work in the particle basis, where individual $N \bar{N}$ channels are treated explicitly.

\subsection{Jacobi coordinates}
\begin{figure}[h]
\centering
\includegraphics[width=0.70\textwidth]{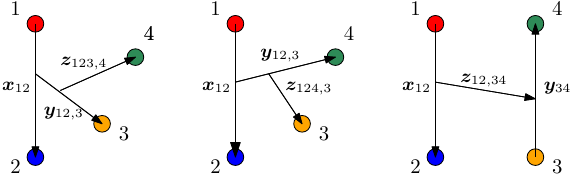}
\caption{Jacobi coordinates for $\mathcal{K}$ and $\mathcal{H}$ partitions including the interaction pair $(12)$.}
\label{fig:jac1234v}
\end{figure}
In this work, Jacobi coordinates are employed to describe four-particle
systems. Two fundamentally different sets of reduced relative coordinates can
be defined, as illustrated in Fig. \ref{fig:jac1234v}. The first set corresponds to the
3+1 particle channels:
\begin{equation}
\begin{tabular}{ll}
$\bm{x}_{ij}$ & $=\sqrt{2\frac{m_{i}m_{j}}{(m_{i}+m_{j})m_{N}}}(%
\bm{r}_{j}-\bm{r}_{i}),\smallskip $ \\
$\bm{y}_{ij,k}$ & $=\sqrt{2\frac{\left( m_{i}+m_{j}\right) m_{k}%
}{(m_{i}+m_{j}+m_{k})m_{N}}}(\bm{r}_{k}-\frac{m_{i}%
\bm{r}_{i}+m_{j}\bm{r}_{j}}{m_{i}+m_{j}}),\smallskip $
\\
$\bm{z}_{ijk,l}$ & $=\sqrt{2\frac{\left(
m_{i}+m_{j}+m_{k}\right) m_{l}}{(m_{i}+m_{j}+m_{k}+m_{l})m_{N}}}(%
\bm{r}_{l}-\frac{m_{i}\bm{r}_{i}+m_{j}%
\bm{r}_{j}+m_{k}\bm{r}_{k}}{m_{i}+m_{j}+m_{k}}%
),\smallskip $%
\end{tabular}%
\   \label{Jacobi_K}
\end{equation}%
where $m_{\ell }$ and $\bm{r}_{\ell}$ denote the mass and the position of the $\ell $-th particle, respectively. The average nucleon mass $m_{N}$ is introduced so that the coordinates $x$, $y$, and $z$ preserve physical distance units.

A second Jacobi coordinate set corresponds to 2+2 particle channels:
\begin{equation}
\begin{tabular}{ll}
$\bm{x}_{ij}$ & $=\sqrt{2\frac{m_{i}m_{j}}{(m_{i}+m_{j})m_{N}}}(%
\bm{r}_{j}-\bm{r}_{i}),\smallskip $ \\
$\bm{y}_{kl}$ & $=\sqrt{2\frac{m_{k}m_{l}}{(m_{k}+m_{l})m_{N}}}(%
\bm{r}_{l}-\bm{r}_{k}),\smallskip $ \\
$\bm{z}_{ij,kl}$ & $=\sqrt{2\frac{\left( m_{i}+m_{j}\right)
\left( m_{k}+m_{l}\right) }{(m_{i}+m_{j}+m_{k}+m_{l})m_{N}}}(\frac{m_{k}%
\bm{r}_{k}+m_{l}\bm{r}_{l}}{m_{k}+m_{l}}-\frac{m_{i}%
\bm{r}_{i}+m_{j}\bm{r}_{j}}{m_{i}+m_{j}})$.%
\end{tabular}%
\   \label{Jacobi_H}
\end{equation}%
The mass-scaled Jacobi coordinates defined above preserve the hyperradius:
\begin{eqnarray}
\rho ^{2} &=&\bm{x}_{ij}^{2}+\bm{y}_{kl}^{2}+%
\bm{z}_{ij,kl}^{2} \\
&=&\bm{x}_{ij}^{2}+\bm{y}_{ij,k}^{2}+\bm{z}_{ijk,l}^{2}.
\end{eqnarray}

\subsection{Faddeev-Yakubovsky equations}
The $\bar{p}+\rm{^3H}$ and $\bar{p}+\rm{^3He}$ systems are treated as systems of four particles interacting through pairwise potentials $\hat{V}^{\alpha,\beta}_{ij}$. Within the optical model framework, the wavefunction consists of two coupled four-particle channels, connected through a charge-exchange interaction term  $\hat{V}^{\alpha,\beta}_{ij}$  with $\alpha\neq\beta$. The wavefunction satisfies the Hamiltonian equation
\begin{equation}
(E-M_\alpha c^2-\hat{H}_{0}) \Psi_\alpha =\sum_{\beta,{(i<j)\in\beta}} \hat{V}_{ij}^{\alpha,\beta}  \, \Psi_\beta, \, \quad \Psi = \begin{pmatrix}
\Psi_{\bar{p}} \\ \Psi_{\bar{n}}
\end{pmatrix},
\end{equation} 
where $E$ is the energy, $H_0$ is the four-body kinetic energy operator, $\Psi_{\bar{p}}$ is the wavefunction of the $\bar{p}+3N$ channel, and the $\Psi_{\bar{n}}$ component corresponds to the  $\bar{n}+3N$ channel. The term $M_\alpha$ denotes the total mass of the four-particle system in channel $\alpha$.  Physical solutions of the Hamiltonian are obtained by deriving and solving modified Faddeev–Yakubovsky (FY) equations, which provide a proper treatment of the Coulomb interaction in configuration space \cite{Y67,LC20}.

The derivation of Faddeev-Yakubovsky equations (FYe's) starts by defining binary
Faddeev components~\cite{Y67}:
\begin{equation}
\varphi^\alpha _{ij}=(E-M_\alpha c^2-\hat{H}_{0})^{-1}\sum_\beta V^{\alpha,\beta}_{ij}\Psi_\beta,
\end{equation}
where $\left( ij\right) $ labels one of the particle pairs. There is one binary Faddeev component for each pairwise interaction. These components vanish when the interaction between particles  $i$ and $j$ goes to zero. For short-range interactions, this property ensures asymptotic decoupling between different components.

The situation changes in the presence of long-range interactions, especially attractive ones.
On one hand, these interactions vanish too slowly to guarantee full asymptotic decoupling of the Faddeev components. On the other hand, in the case of the attractive Coulomb interaction, the short-range part is very strong—indeed divergent—and is responsible for binding the particles within the clusters formed during the collision. To overcome this difficulty in three-body systems, Merkuriev
proposed a solution~\cite{M80,GRY16}, based
on the splitting of the Coulomb interaction into short- and long-range parts:
\begin{equation}
{V}^C_{ij}={V}_{ij}^{sh}+{V}_{ij}^{lo}.
\end{equation}
According to Merkuriev’s ansatz, the derivation of the Faddeev–Yakubovsky (FY) equations for charged particles proceeds by replacing the free Hamiltonian with an effective Hamiltonian that incorporates all long-range interaction terms:
$\hat{H}_0\rightarrow \hat{H}^\alpha_{0}=\hat{H}_{0}-M_\alpha c^2+\sum\limits_{(k<l)\in\alpha}{V}_{kl}^{lo}$. 
The long-range contributions originate solely from the Coulomb interaction and contain no charge-exchange terms; they are therefore diagonal in the channel index $\alpha$. In this way, each full pair interaction $V_{ij}$ is replaced by a short-range potential $\upsilon_{ij}$, obtained by subtracting the long-range Coulomb component.
Following the iterative procedure of Yakubovsky~\cite{Y67}, the binary Faddeev components are redefined as
\begin{equation}
\tilde{\varphi} _{ij}^{\alpha}=(E-\hat{H}_{0}^\alpha)^{-1}\sum_\beta\hat{\upsilon}_{ij}^{\alpha,\beta}\Psi_\beta.
\end{equation}
As with the standard Faddeev components, one easily verifies that the total four-body wavefunction can be written as a sum of the newly defined Faddeev–Merkuriev components ($\tilde{\varphi}_{ij}$):
\begin{equation}
\Psi^\alpha =\sum\limits_{i<j}\varphi _{ij}^\alpha=\sum\limits_{i<j}\tilde{\varphi} _{ij}^{\alpha}.
\end{equation}

To guarantee asymptotic decoupling of the binary components in the collision channels, it is essential to satisfy Merkuriev’s criteria for the Coulomb-splitting procedure.
Specifically, the short-range potential $v_{ij}$ must fulfil two key conditions: 
\begin{enumerate}
  \item It must remain bounded and vanish rapidly enough to satisfy the Faddeev condition
  \begin{equation}
\upsilon_{ij}(x_{ij},y_{ij,k})=0;\text{ for }%
x_{ij}>a_{0}(1+y_{ij,k})^{{1/\nu\prime} },
\end{equation}
where $a_0$ is some constant and $\nu ^{\prime }$ is a parameter larger than 2.
  \item A complementary requirement ensures that the modified channel Hamiltonians $(E-H^\alpha_{0}) $ do not generate spurious asymptotic bound states. As shown by Merkuriev, this condition leads to 
  \begin{equation}
\upsilon_{ij}(x_{ij},y_{ij,k})=V_{ij}(x_{ij});\text{ for }%
x_{ij}<a_{0}(1+y_{ij,k})^{1/\nu },
\end{equation}
where $2<\nu ^{\prime }<\nu.$
\end{enumerate}

If the Coulomb splitting is expressed as a function of the hyperradius $\rho _{ij}=\sqrt{\rho ^{2}-x_{ij}^{2}}$ , then the Yakubovsky procedure can be followed directly to define the Faddeev–Yakubovsky components
 $\mathcal{K}_{ij,k}^{l},$ associated with $3+1$ channels, and $%
\mathcal{H}_{ij}^{kl},$ associated with $2+2$ channels:
\begin{eqnarray}
{^\alpha \mathcal{K}}_{ij,k}^{l} &=&(E-\hat{H}_{0}^\alpha)^{-1}\sum_\beta\hat{\upsilon}^{\alpha,\beta}_{ij}(\tilde{\varphi} _{ik}^\beta+\tilde{\varphi}^\beta_{jk}+{^\beta \mathcal{K}}_{ij,k}^{l}),
\label{FYMK-eqs} \\
{^\alpha \mathcal{H}}_{ij}^{kl} &=&(E-\hat{H}_{0}^\alpha)^{-1}\sum_\beta\hat{\upsilon}_{ij}^{\alpha,\beta}(\tilde{\varphi} _{kl}\beta_{jk}+{^\beta \mathcal{H}}_{ij}^{kl}).  \label{FYMH-eqs}
\end{eqnarray}%
In a four-body system, there are twelve  $K$-type components and six
 $H$-type components, obtained by permuting the particle indices $\left(
ijkl\right)$. Due to the presence of two distinct particle channels, we have a
total of 2$\times$18 components, which satisfy a coupled system of 36 differential equations.
 The 3-body FM components are then reconstructed as
\begin{equation}
\tilde{\varphi}  _{ij}^\alpha={^\alpha \mathcal{K}_{ij,k}^{l}}+{^\alpha \mathcal{K}_{ij,l}^{k}}+{^\alpha \mathcal{H}_{ij}^{kl}},
\end{equation}
and the total wavefunction follows from summing over all FY components:
\begin{equation}
\Psi^\alpha =\sum_{(i<j,k,l)=1}^{4}{^\alpha \mathcal{K}}_{ij,k}^{l}+\sum_{(i<j,k<l)=1}^{4}{^\alpha\mathcal{H}}_{ij}^{kl}.
\end{equation}%
Although the formal four-body equations satisfy Merkuriev’s criteria and guarantee the decoupling of binary-channel asymptotes, it is advantageous—especially for numerical implementations—to minimize the overlap between distinct FY components and enforce rapid and smooth decoupling in asymptotic regions.

To implement physical scattering boundary conditions, the FY equations must reduce to equations for 3+1 or 2+2 bound subclusters as the respective separations z$_{ijl,k}$ or z$_{ij,kl}$ become large. This motivates the use of distinc Coulomb-splitting functions for $2+2$ and $3+1$ channels. In this context, three functionals for splitting the Coulomb interaction are introduced:
\begin{eqnarray}
V^C_{a}(x_{a}) &=&V_{a}^{sh_{K}}(x_{a},y_{a})+V_{a}^{lo_{K}}(x_{a},y_{a}),\qquad
a\in K  \label{split_a} \\
V^C_{b}(x_{b}) &=&V_{b}^{sh_{H}}(x_{b},y_{b})+V_{b}^{lo_{H}}(x_{b},y_{b}),\qquad
b\in H  \label{split_b} \\
V^C_{c}(x_{c}) &=&V_{c}^{sh_{\rho }}(x_{c},\rho _{c})+V_{c}^{lo_{\rho
}}(x_{c},\rho _{c}),\qquad c\in K,H  \label{split_c} \\
\rho _{c} &=&\sqrt{y_{c}^{2}+z_{c}^{2}}.
\end{eqnarray}%
The standard splitting function proposed in Ref.~\cite{M80} is employed:
\begin{equation}
V^{sh}(x,y)=V^{C}(x)-V^{lo}(x,y)=\frac{2}{1+\exp [\frac{\left( x/x_{0}\right)
^{\nu }}{1+y/y_{0}}]}V^{C}(x).
\end{equation}

The resulting modification of the four-body FY equations reads:
\begin{eqnarray}
(E-\hat{H}^\alpha_{0}-V_{ij}^{lo_{K}}-V_{ik}^{lo_{K}}-V_{jk}^{lo_{K}}-V_{il}^{lo_{\rho
}}-V_{jl}^{lo_{\rho }}-V_{kl}^{lo_{\rho }})^\alpha\mathcal{K}_{ij,k}^{l} &=&\sum_\beta\upsilon_{ij}^{{\alpha,\beta}_{\rho
}}(\tilde{\varphi}^\beta _{ik}+^\beta\tilde{\varphi} _{jk}
^\beta\mathcal{K}_{ij,k}^{l})\notag\\
+(V_{ij}^{sh_{K}}-V_{ij}^{sh_{\rho
}})(^\alpha\mathcal{K}_{ij,k}^{l}+^\alpha\mathcal{K}_{ik,j}^{l}+^\alpha\mathcal{K}_{jk,i}^{l}), \\
(E-\hat{H}^\alpha_{0}-V_{ij}^{lo_{H }}-V_{kl}^{lo_{H}}-V_{ik}^{lo_{\rho }}-V_{jk}^{lo_{\rho }}-V_{il}^{lo_{\rho
}}-V_{jl}^{lo_{\rho }})^\alpha\mathcal{H}_{ij}^{kl} &=&\sum_\beta\upsilon_{ij}^{{\alpha,\beta}_{\rho
}}(^\beta\tilde{\varphi}_{kl}+^\beta\mathcal{H}_{ij}^{kl})\notag\\
+(V_{ij}^{sh_{H}}-V_{ij}^{sh_{\rho }})(^\alpha\mathcal{H}_{ij}^{kl}+^\alpha\mathcal{H}_{kl}^{ij}).
\end{eqnarray}

By permuting particle indices, one obtains a system of 2$\times$18 differential equations describing the scattering problem with two coupled four-body channels, each containing four distinct particles. When symmetries under the exchange of identical particles are taken into account, several FY components can be expressed in terms of others, thereby reducing the number of independent equations. Table \ref{tab:FY} summarizes the particle channels and associated FY components for the $\bar{p}+\rm{^3H}$ and $\bar{p}+\rm{^3He}$ systems.
 \begin{table}[h]
 \small
    \centering
    \begin{tabular}{ccccccccccccc}
    \colrule
$\bar{p}+\rm{^3H}$ & $p \bar{p} n n$ & $\mathcal{K}^{n}_{p \bar{p},n}$ & $\mathcal{K}^{n}_{p n,\bar{p}}$ & $\mathcal{K}^{\bar{p}}_{p n,n}$ & $\mathcal{K}^{p}_{\bar{p}n,n}$ & $\mathcal{K}^{n}_{\bar{p}n,p}$ & $\mathcal{K}^{p}_{nn,\bar{p}}$ & $\mathcal{K}^{\bar{p}}_{nn,p}$ & $\mathcal{H}_{p \bar{p},nn}$ & $\mathcal{H}_{pn,\bar{p}n}$ & $\mathcal{H}_{\bar{p}n,pn}$ & $\mathcal{H}_{nn,p\bar{p}}$  \\ 
& $n \bar{n} n n$ & $\mathcal{K}^{n}_{n \bar{n},n}$ & $\mathcal{K}^{n}_{nn,\bar{n}}$ & $\mathcal{K}^{\bar{n}}_{n n,n}$ & $\mathcal{H}_{n \bar{n},nn}$ & $\mathcal{H}_{n n,\bar{n}n}$   \\ 
\hline
$\bar{p}+\rm{^3He}$ & $n \bar{p} p p$ & $\mathcal{K}^{p}_{n \bar{p},p}$ & $\mathcal{K}^{p}_{n p,\bar{p}}$ & $\mathcal{K}^{\bar{p}}_{np,p}$ & $\mathcal{K}^{n}_{\bar{p}p,p}$ & $\mathcal{K}^{p}_{\bar{p}p,n}$ & $\mathcal{K}^{n}_{pp,\bar{p}}$ & $\mathcal{K}^{\bar{p}}_{pp,n}$ & $\mathcal{H}_{n \bar{p},pp}$ & $\mathcal{H}_{np,\bar{p}p}$ & $\mathcal{H}_{\bar{p}p,np}$ & $\mathcal{H}_{pp,n\bar{p}}$  \\ 
& $n \bar{n} n p$ &  $\mathcal{K}^{n}_{p \bar{n},n}$ & $\mathcal{K}^{n}_{p n,\bar{n}}$ & $\mathcal{K}^{\bar{n}}_{p n,n}$ & $\mathcal{K}^{p}_{\bar{n}n,n}$ & $\mathcal{K}^{n}_{\bar{n}n,p}$ & $\mathcal{K}^{p}_{nn,\bar{n}}$ & $\mathcal{K}^{\bar{n}}_{nn,p}$ & $\mathcal{H}_{p \bar{n},nn}$ & $\mathcal{H}_{pn,\bar{n}n}$ & $\mathcal{H}_{\bar{n}n,pn}$ & $\mathcal{H}_{nn,p\bar{n}}$ \\
\colrule
\end{tabular}
\caption{Four-body particle channels and associated FY components for $\bar{p}+\rm{^3H}$ and $\bar{p}+\rm{^3He}$ problems.}
\label{tab:FY}
\normalsize
\end{table}

\subsection{Numerical resolution}
For a quantum state characterised by the total angular momentum $J$ and the parity $\Pi$, the FY components are expanded in partial waves as
\begin{equation}
\mathcal{F}(\bm{x},\bm{y},\bm{z}) = \sum_{n} \frac{F_n(x,y,z)}{xyz} \, \mathcal{Y}^{(F)}_{n}(\hat{x},\hat{y},\hat{z}),
\end{equation}
where $F_n$ is the radial component of partial wave $n$ and $\mathcal{Y}_n$ denotes tripolar spherical harmonics involving the coupling of orbital angular momenta and spins. For the
K- and H-type components, respectively:
\begin{align}
&\mathcal{Y}^{(K)}_{n} = \left[ \left[ \left[l_x (s_i s_j)_{s_x}\right]_{j_x} (l_y s_k)_{j_y} \right]_{j_{xy}} (l_z s_l)_{j_z} \right]_{J^{\pi}}, \\
&\mathcal{Y}^{(H)}_{n} = \left[ \left[ \left[l_x (s_i s_j)_{s_x}\right]_{j_x} \left[l_y (s_k s_l)_{s_y}\right]_{j_y} \right]_{j_{xy}} l_z \right]_{J^{\pi}}.
\end{align}
The index $n$ represents a set of eight quantum numbers defining the overall angular momentum and parity, and including the orbital momenta associated with each Jacobi coordinate as well as quantum numbers representing partial couplings with individual spins. The radial functions are expanded over regularized Lagrange–Laguerre bases \cite{B15}:
\begin{equation}
F_n(x,y,z) = \sum_{i_x=1}^{k_x} \sum_{i_y=1}^{k_y} \sum_{i_z=1}^{k_z} c^{(F)}_{n i_x i_y i_z} \, \hat{f}_{i_x}\left(\frac{x}{h_x} \right) \, \hat{f}_{i_y}\left(\frac{y}{h_y} \right) \, \hat{f}_{i_z}\left(\frac{z}{h_z} \right),
\end{equation}
where the coefficients $c^{(F)}_{n i_x i_y i_z}$ are determined by solving a generalized eigenvalue problem for bound states and linear systems for scattering states. The scaling parameters $h_x$, $h_y$, and $h_z$ adjust the spatial extension of the grid for each Jacobi coordinate to the range of the interaction; in the present work, we use $h_x=h_y=h_z=0.2-0.3 \, \text{fm}$. Convergence is verified by increasing both the number of partial waves and the number of grid points along each Jacobi coordinate. The results presented below were obtained with $k=k_x=k_y=k_z=20-25$ and by considering all partial waves such that $l_x+l_y+l_z \leq 4$, which was found sufficient for convergence. All calculations use the average nucleon mass $m_N =~938.94 \, \frac{\text{MeV}}{c^2}$.

\section{Results}
\subsection{Scattering lengths}
The first step in the evaluation of the atomic level shifts is the calculation of the scattering lengths for the $\bar{p}+\rm{^3H}$ and $\bar{p}+\rm{^3He}$ systems. We adopt the Kohno–Weise (KW) potential as the reference $N \bar{N}$ interaction and assess the model dependence by comparing results obtained with different  $NN$  potentials: the phenomenological MT-I-III \cite{MT69}, the semi-realistic AV18 meson-exchange interaction \cite{AV18} and the chiral effective field theory–based I-N3LO \cite{MN3LO}  potential.

The calculated scattering lengths for the $S$-wave ($J^{\Pi}=0^+$, $1^+$) and $P$-wave ($J^{\Pi}=0^-$, $1^-$, $2^-$) states are presented in Tables \ref{tab:a_NN_3H} and \ref{tab:a_NN_3He}. For the $J^{\Pi}=1^-$ channel, the scattering wavefunction involves two asymptotic partial waves ($l_z=1$, $j_{z}=\frac{1}{2}$ and $l_z=1$, $j_z=\frac{3}{2}$), yielding a $2 \times 2$ matrix. For both $\bar{p}+\rm{^3H}$ and $\bar{p}+\rm{^3He}$, the calculated scattering lengths show only a very weak dependence on the underlying $NN$ interaction. This is consistent with our previous results for antiproton–deuteron scattering~\cite{CL21,*DLC23}. 

Although a fully realistic description of $\rm{^3H}$ and $\rm{^3He}$ nuclei would include three-nucleon forces, the close agreement between the results obtained with the central MT-I-III potential and those from realistic interactions -- despite their $\sim10 \%$ differences in three-body binding energies -- indicates that three-nucleon force contribution is negligible in the present context. This insensitivity reflects the peripheral character of the antiproton annihilation process, which depends only weakly on the off-shell structure of the strong interaction. It also suggests that possible contributions of $\bar{N}NN$ forces are unlikely to play a significant role in the low-energy observables considered here.

\begin{table}[h]
    \centering
    \begin{tabular}{cccc} \toprule
 $\bar{p}+\rm{^3H}$ & MT-I-III+KW  & AV18+KW & I-N3LO+KW \\ \colrule

  $S$ states & \multicolumn{3}{c}{$a_0 \, (\text{fm})$} \\
  \colrule
  $0^+$   & $1.42-0.75 \, $i  & $1.44-0.76 \, $i & $1.44-0.77 \, $i\\  
  $1^+$   & $1.59-0.65 \, $i  & $1.59-0.65 \, $i & $1.60-0.65 \, $i\\
  \colrule
  $P$ states &  \multicolumn{3}{c}{$a_1 \, (\text{fm}^3)$}\\
  \colrule
  $0^-$   & $-1.09-4.45 \, $i  & $-0.84-4.47 \, $i & $-0.78-4.63 \, $i\\
  $1^-$   & $\begin{pmatrix} 1.67-2.84 \, \rm{i} & 1.28-0.34\, \rm{i} \\ 1.28-0.34 \, \rm{i} & 0.78-3.47 \, \rm{i} \end{pmatrix}$ & $\begin{pmatrix} 1.75-2.95 \, \rm{i} & 1.39-0.33\, \rm{i} \\ 1.39-0.33 \, \rm{i} & 0.79-3.62 \, \rm{i} \end{pmatrix}$ & $\begin{pmatrix} 1.85-2.95 \, \rm{i} & -1.50-0.32 \, \rm{i} \\ -1.50-0.32 \, \rm{i} & 0.82-3.60 \, \rm{i} \end{pmatrix}$ \\
  $2^-$   & $1.35-3.35 \, $i  & $1.23-3.35 \, $i & $ 1.24-3.36 \, $i\\
  \toprule
    \end{tabular}
    \caption{Scattering lengths of $\bar{p}+\rm{^3H}$ $S$ and $P$ states computed with different $NN$ interactions, used in conjunction with the KW $N \bar{N}$ potential.}
    \label{tab:a_NN_3H}
\end{table}

\begin{table}[h]
    \centering
    \begin{tabular}{cccc} \toprule
 $\bar{p}+\rm{^3He}$ & MT-I-III+KW  & AV18+KW & I-N3LO+KW \\ \colrule

  $S$ states & \multicolumn{3}{c}{$a_0 \, (\text{fm})$} \\
  \colrule
  $0^+$   & $1.49-0.58 \, $i  & $1.50-0.58 \, $i & $1.50- 0.62 \, $i\\  
  $1^+$    & $1.42-0.67 \, $i   & $1.42-0.67 \, $i & $1.44- 0.69 \, $i\\
  \colrule
  $P$ states &  \multicolumn{3}{c}{$a_1 \, (\text{fm}^3)$}\\
  \colrule
  $0^-$  & $2.74-2.52 \, $i & $3.02-2.78 \, $i & $2.99-2.73 \, $i\\
  $1^-$    & $\begin{pmatrix} 0.34-3.49 \, \rm{i} & 1.23+0.58 \, \rm{i} \\ 1.23+0.58 \, \rm{i} & 1.13-4.00 \, \rm{i} \end{pmatrix}$  & $\begin{pmatrix} 0.44-3.51 \, \rm{i} & 1.19+0.56 \, \rm{i} \\ 1.19+0.56 \, \rm{i} & 1.27-4.06 \, \rm{i} \end{pmatrix}$ & $\begin{pmatrix} 0.74- 3.75 \, \rm{i} & 1.28+ 0.61 \, \rm{i} \\ 1.28 + 0.61 \, \rm{i} &  1.46- 4.48 \, \rm{i} \end{pmatrix}$ \\
  $2^-$   &  $1.02-3.18 \, $ i   & $1.10-3.29 \, $i & $1.12-3.23 \, $i\\
  \toprule
    \end{tabular}
    \caption{Scattering lengths of $\bar{p}+\rm{^3He}$ $S$ and $P$ states computed with different $NN$ interactions, used in conjunction with the KW $N \bar{N}$ potential.}
    \label{tab:a_NN_3He}
\end{table}
Using the I-N3LO interaction as the reference 
$NN$ potential, we next compare scattering lengths obtained with two different $N \bar{N}$ models. Tables \ref{tab:a_NNbar_3H} and \ref{tab:a_NNbar_3He} present the results obtained with the KW and Jülich interactions for the $\bar{p}+\rm{^3H}$ and $\bar{p}+\rm{^3He}$ systems, respectively. For the $S$-wave states, the two
 $N \bar{N}$ models show good agreement despite their markedly different theoretical formulations and parametrizations: both the real and imaginary parts of the scattering lengths differ by less than $10\%$. In contrast, the $P$-wave states exhibit pronounced model dependence in both systems. The largest discrepancies appear in the real part of the scattering lengths, reaching up to an order of magnitude, while the imaginary parts are comparatively more stable. A similar behaviour was previously observed in the $P$ states of the $\bar{p}d$ system and was attributed to the limited empirical constraints on the $N \bar{N}$ interaction, particularly for partial waves with $l>0$. Due to the lack of low-energy experimental data, existing $N \bar{N}$ potential models can differ substantially, leading to sizeable discrepancies not only in few-body systems but also at the two-body level~\cite{CHW23}.

\begin{table}[h]
    \centering
    \begin{tabular}{ccc} \toprule
 $\bar{p}+\rm{^3H}$ & I-N3LO+KW  & I-N3LO+J\"ulich \\ \colrule
  $S$ states & \multicolumn{2}{c}{$a_0 \, (\text{fm})$} \\
  \colrule
  $0^+$   & $1.44-0.77 \, $i  & $ 1.28-0.80 \, $i\\  
  $1^+$   & $1.60-0.65 \, $i  & $1.45-0.83 \, $i \\
  \colrule
  $P$ states &  \multicolumn{2}{c}{$a_1 \, (\text{fm}^3)$}\\
  \colrule
  $0^-$   & $-0.78-4.63 \, $i  & $-1.58-5.69 \, $i \\
  $1^-$   & $\begin{pmatrix} 1.85-2.95 \, \rm{i} & -1.50-0.32 \, \rm{i} \\ -1.50-0.32 \, \rm{i} & 0.82-3.60 \, \rm{i} \end{pmatrix}$ & $\begin{pmatrix} 1.21-2.46 \, \rm{i} & -1.3- 0.18 \, \rm{i} \\ -1.3- 0.18\, \rm{i} & 0.12 - 3.00 \, \rm{i} \end{pmatrix}$  \\
  $2^-$   & $1.24-3.36 \, $i  & $ 0.14-3.15 \, $i \\
  \toprule
    \end{tabular}
    \caption{Scattering lengths of $\bar{p}+\rm{^3H}$ $S$ and $P$ states computed with different $N \bar{N}$ interactions, used in conjunction with the I-N3LO $NN$ potential.}
    \label{tab:a_NNbar_3H}
\end{table}

\begin{table}[h]
    \centering
    \begin{tabular}{ccc} \toprule
 $\bar{p}+\rm{^3He}$ & I-N3LO+KW  & I-N3LO+J\"ulich \\ \colrule
  $S$ states & \multicolumn{2}{c}{$a_0 \, (\text{fm})$} \\
  \colrule
  $0^+$   & $1.50- 0.62 \, $i  & $1.39- 0.69 \, $i\\  
  $1^+$   & $1.44- 0.69 \, $i  & $1.34- 0.90 \, $i \\
  \colrule
  $P$ states &  \multicolumn{2}{c}{$a_1 \, (\text{fm}^3)$}\\
  \colrule
  $0^-$   & $2.99-2.73 \, $i  & $ 2.54- 3.21 \, $i \\
  $1^-$   & $\begin{pmatrix} 0.74- 3.75 \, \rm{i} & 1.28+ 0.6 \, \rm{i} \\ 1.28 + 0.6 \, \rm{i} &  1.46- 4.48 \, \rm{i} \end{pmatrix}$ & $\begin{pmatrix} 0.42 - 4.16 \, \rm{i} & 1.5+ 0.75 \, \rm{i} \\ 1.5 + 0.75 \, \rm{i} &  0.46-4.31 \, \rm{i} \end{pmatrix}$  \\
  $2^-$   & $1.12-3.23 \, $i  & $ 0.11- 3.26\, $i \\
  \toprule
    \end{tabular}
    \caption{Scattering lengths of $\bar{p}+\rm{^3He}$ $S$ and $P$ states computed with different $N \bar{N}$ interactions, used in conjunction with the I-N3LO $NN$ potential.}
    \label{tab:a_NNbar_3He}
\end{table}

\subsection{Level shifts}
Using the scattering lengths listed in Tables \ref{tab:a_NNbar_3H} and \ref{tab:a_NNbar_3He}, the level shifts of the low-lying hydrogenic states of $\bar{p} \, \rm{^3H}$ and $\bar{p} \, \rm{^3He}$ systems are obtained by means of the Trueman formula \eqref{Tr}. The results obtained with the KW and J\"ulich $N \bar{N}$ potentials are listed in Tables \ref{tab:E_NNbar_3H} and \ref{tab:E_NNbar_3He}. Since the level shifts follow directly from the corresponding scattering lengths, the same observations regarding model dependence apply.

Because the $\bar{p} \, \rm{^3He}$ system has nuclear charge $Z=2$, its Coulomb energy levels scale as $Z^2$, i.e. they are a factor of four larger than those of $\bar{p} \, \rm{^3H}$, while the corresponding Bohr radius is reduced by a factor of two. Consequently, the level shifts in $\bar{p} \, \rm{^3He}$ are typically an order of magnitude larger than those in $\bar{p} \, \rm{^3H}$. It should also be noted that the Trueman formula is constructed as a series expansion in $\frac{a_l}{B}$. For systems with larger nuclear charge (and hence smaller Bohr radii), higher-order terms may be required to achieve quantitatively accurate predictions.

\begin{table}[h]
    \centering
    \begin{tabular}{ccc} \toprule
 $\bar{p} \, \rm{^3H}$ & I-N3LO+KW  & I-N3LO+J\"ulich \\ \colrule
  $S$ states & \multicolumn{2}{c}{$\Delta E_{ns} \, (\text{keV})$} \\
  \colrule
  $0^+$ ($n=1$)  & $2.57-1.15 \, $i  & $2.33-1.23 \, $i\\  
  $1^+$ ($n=1$)  & $2.78-0.94 \, $i  & $2.61-1.23 \, $i \\
  \colrule
  $P$ states &  \multicolumn{2}{c}{$ \Delta E_{pn} \, (\text{meV})$}\\
  \colrule
  $0^-$ ($n=2$)   & $-96.8-575 \, $i  & $-196-706 \, $i \\
  $1^-$ ($n=2$)   & $361-356 \, $i & $256-305 \, $i   \\
                  & $-29.6-458 \, $i &  $-91.1-373 \, $i  \\
  $2^-$ ($n=2$)  & $154-418 \, $i  & $17.3-391 \, $i \\
  \toprule
    \end{tabular}
    \caption{Level shifts of $\bar{p} \, \rm{^3H}$ $S$ and $P$ states computed with different $N \bar{N}$ interactions, used in conjunction with the I-N3LO $NN$ potential.}
    \label{tab:E_NNbar_3H}
\end{table}

\begin{table}[h]
    \centering
    \begin{tabular}{ccc} \toprule
 $\bar{p} \, \rm{^3He}$ & I-N3LO+KW  & I-N3LO+J\"ulich \\ \colrule
  $S$ states & \multicolumn{2}{c}{$\Delta E_{ns} \, (\text{keV})$} \\
  \colrule
  $0^+$ ($n=1$)  & $18.6- 4.87 \, $i  & $18.0- 5.86 \, $i\\  
  $1^+$ ($n=1$)  & $  18.4 - 5.68
 \, $i  & $ 18.4- 7.83 \, $i \\
  \colrule
  $P$ states &  \multicolumn{2}{c}{$ \Delta E_{pn} \, (\text{eV})$}\\
  \colrule
  $0^-$ ($n=2$)   & $11.9-10.8 \, $i  & $10.1 -12.8 \, $i \\
  $1^-$ ($n=2$)   & $-0.55-18.4 \, $i & $-4.19 -19.8 \, $i   \\
                  & $ 9.30-14.3 \, $i &  $7.71 -13.8 \, $i  \\
  $2^-$ ($n=2$)  & $4.45-12.8 \, $i  & $ 0.44 -13.0 \, $i \\
  \toprule
    \end{tabular}
    \caption{Level shifts of $\bar{p} \, \rm{^3He}$ $S$ and $P$ states computed with different $N \bar{N}$ interactions, used in conjunction with the I-N3LO $NN$ potential.}
    \label{tab:E_NNbar_3He}
\end{table}

Experimentally, the $\bar{p} \, \rm{^3He}$ system has been studied using data collected at the Low Energy Antiproton Ring (LEAR) \cite{DG84_pbarHe,SB91_3he_4he,BB89_pbarHe,B90_lighpbar_he_k,BG00_pb3He}. Tritium was not investigated due to radiological safety considerations. Although individual hyperfine levels could not be resolved, spin-averaged level shifts of the low-lying $P$ states were extracted. The measured $2p$ shift \cite{SB91_3he_4he} is given in Table \ref{tab:Ecomp}, together with our \textit{ab initio} predictions obtained using different $N \bar{N}$ models, as well as earlier theoretical results \cite{Schneider_PhD,B90_lighpbar_he_k,BW20}. 

The results reported in Refs. \cite{Schneider_PhD, B90_lighpbar_he_k} employ effective antiproton–nucleus potentials of the optical form
\begin{equation}
U(r) = -\left(\frac{2 \pi}{\mu_{\bar{p}A}} \right) \left(1+\frac{\mu_{\bar{p}A}}{m_N} \right) \, \bar{a} \, \rho(r), \label{V_rho}
\end{equation}
where $\mu_{\bar{p}A}$ is the reduced mass, 
$m_N$  the nucleon mass, $\bar{a}$ a complex effective scattering length, and $\rho(r)$ the nuclear density. In this approach, few-body correlations are neglected and the parameters of $\bar{a}$ were adjusted through a global fit to level-shift data for nuclei with $Z \geq 2$. Consequently, such models have limited predictive power and mainly serve as a test of the commonly used ansatz \eqref{V_rho} \cite{ES70,DL74_Urho,BD72_Urho} for exotic atoms. 

In Ref. \cite{BW20}, the level shifts are computed within an effective three-body model consisting of an antiproton, a nucleon, and a residual nucleus. The shifts are expressed in terms of $S$- and $P$-wave scattering amplitudes derived from the Paris $N \bar{N}$ potential. While this framework is not fully microscopic, it relies on a well-established $N \bar{N}$ interaction rather than on an empirical density-dependent potential. The predicted widths agree reasonably well with experiment. However, the Paris 09 model fails to reproduce the repulsive contribution to the $2p$ level, leading to an underestimation of the real part of the shift.

The present work provides the first \textit{ab initio} predictions that simultaneously employ realistic $NN$ and $N \bar{N}$ interactions. As shown in Table~\ref{tab:Ecomp}, both KW and J\"ulich models yield a $2p$ width $\bar{\Gamma}_{2p}$ consistent with the experimental value, well within the associated uncertainties. This result contrasts with the $\bar{p}d$ system, for which neither microscopic calculations nor effective models are able to reproduce the observed $2p$ width \cite{CL21,*DLC23}. For the real part of the energy shift, effective models typically underestimate the $2p$ shift \cite{SB91_3he_4he} and may differ substantially from one another, indicating strong sensitivity to model assumptions. A similar trend emerges in our \textit{ab initio} results, where both considered interactions yield shifts that are approximately an order of magnitude smaller than the experimental value.

\begin{table}[h]
\centering
\begin{tabular}{cccccc} \colrule
& $\overline{\Delta E_R}_{2p}$ (eV) & $\overline{\Gamma}_{2p}$ (eV) \\ \colrule
\textbf{This work} &  \\
\hline
I-N3LO+KW &  $5.04$ & $28.8$  \\
I-N3LO+J\"ulich & $1.91$ & $29.8$  \\
\hline
\textbf{Literature} &  \\
\hline
Schneider, \textit{et al.} \cite{Schneider_PhD} (Opt. potential fit) & $16$ & $21$  \\
Batty \textit{et al.} \cite{B90_lighpbar_he_k} (Opt. potential fit) & $6(1)$ & $24(2)$  \\
Loiseau \textit{et al.} \cite{BW20} (Paris 99) & $12.59$ & $29.8$ \\
Loiseau \textit{et al.} \cite{BW20} (Paris 09) & $1.46$ & $31.4$ \\
\hline
\textbf{Exp.} &  \\
\hline
Ref. \cite{SB91_3he_4he} & $17(4)$ & $25(9)$   \\
\colrule
\end{tabular}
\caption{Spin-averaged level shift of the $\bar{p} \, \rm{^3He}$ $2p$ state computed with different $N \bar{N}$ optical potentials. The values are compared with results published in the literature \cite{Schneider_PhD,B90_lighpbar_he_k,BW20} and experimental data \cite{SB91_3he_4he}.}
\label{tab:Ecomp}
\end{table}

\subsection{Annihilation densities}
The PUMA experiment aims to probe the tails of proton and neutron densities by exploiting antiproton annihilation on nucleons, which is assumed to occur predominantly in the nuclear periphery. This central hypothesis can be tested quantitatively by employing full four-body wavefunctions to compute the annihilation densities associated with low-lying hydrogenic states \cite{S87,CIR89,H89}. The width of a given state $\alpha \equiv J^{\Pi}$ can directly be deduced from the expectation value of the Hamiltonian:
\begin{equation}
    \Gamma = -2 \, \text{Im}\left[\mel{\Psi_{\alpha}}{H}{\Psi_{\alpha}} \right].
\label{Gamma}
\end{equation}
Equation \eqref{Gamma} can be recast as a one-dimensional integral,
\begin{equation}
    \Gamma = -2 \int_{0}^{\infty} \gamma^{(\alpha)}_{a}(z) \, \mathrm{d} z,
\end{equation}
where $\gamma^{(\alpha)}_{a}$ denotes the annihilation density of state $\alpha$ and $z$ is the Jacobi coordinate representing the distance between the antiproton and the nuclear center of mass. Within the optical-model framework, the annihilation density is expressed as
\begin{equation}
    \gamma^{(\alpha)}_{a}(z) = \int |\Psi_{\alpha}(\bm{x},\bm{y},\bm{z})|^2 \, \text{Im}\left[W(x,y,z) \right] \,  x^2 y^2 z^2 \, \mathrm{d} \bm{x} \, \mathrm{d} \bm{y} \, \mathrm{d} \hat{z},
\end{equation}
where $W$ is the absorptive part of the potential and the integration extends over all coordinates except the radial variable $z$. 

The function $\gamma^{(\alpha)}_{a}(z)$ may be interpreted as the probability density for the annihilation process. By comparing it with the radial density of the target nucleus, one can identify the spatial regions where annihilation is most likely to occur, thereby directly testing the key assumption underlying the PUMA program. In the present work, the annihilation densities are derived from the zero-energy scattering wavefunction, which is expected to provide an accurate approximation of the corresponding hydrogenic-state wavefunction in the region relevant for annihilation. 

\begin{figure}[h]
\centering
\includegraphics[width=1.0\textwidth]{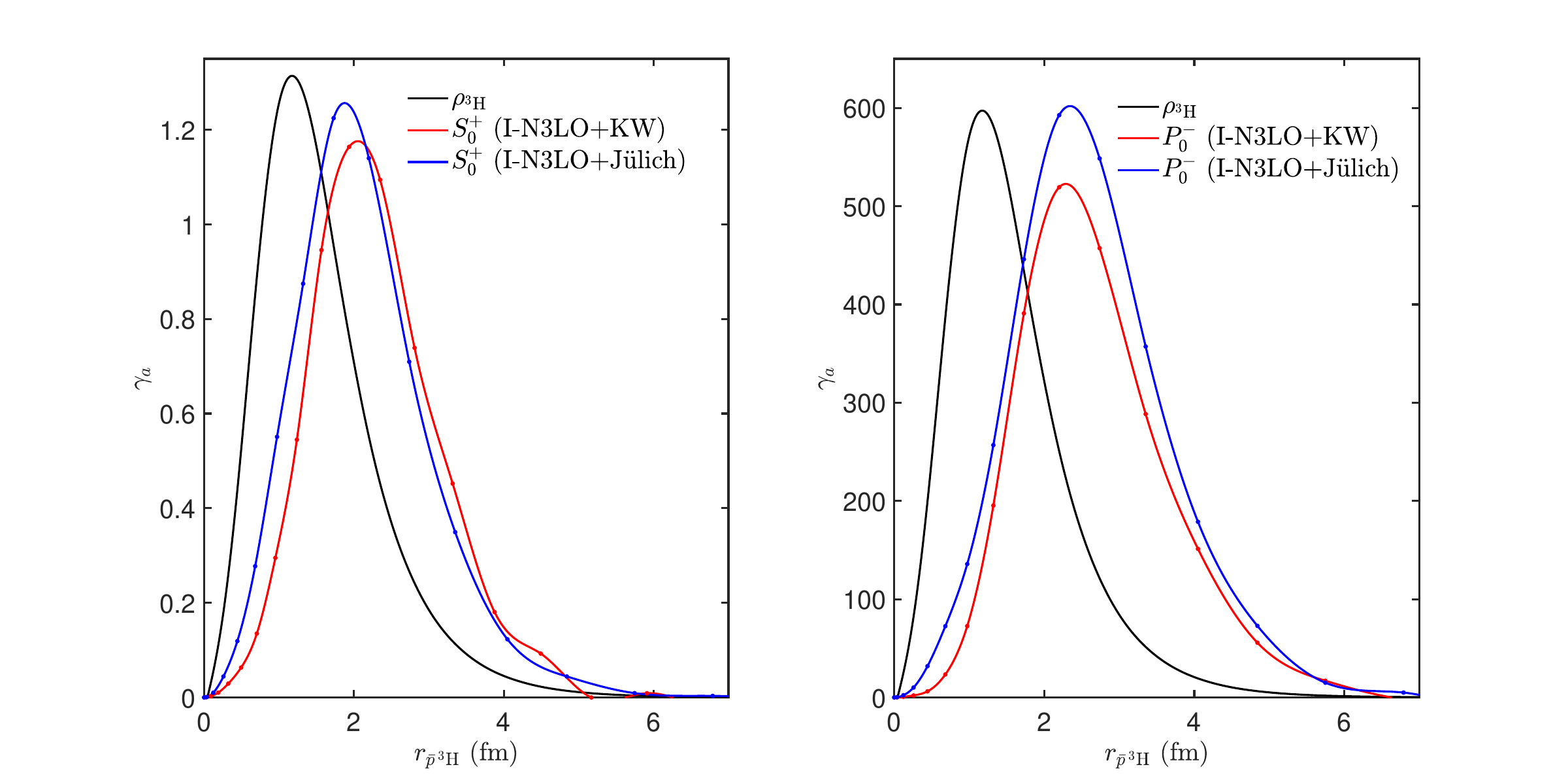}
\caption{Annihilation densities of $0^+$ (left) and $0^-$ (right) $\bar{p} \, \rm{^3H}$ states computed with the KW (red) and J\"ulich (blue) potentials, used in conjunction with the I-N3LO interaction. The curves are compared with the appropriately scaled density of $\rm{^3H}$ (black).}
\label{fig:g3H}
\end{figure}

\begin{figure}[h]
\centering
\includegraphics[width=1.0\textwidth]{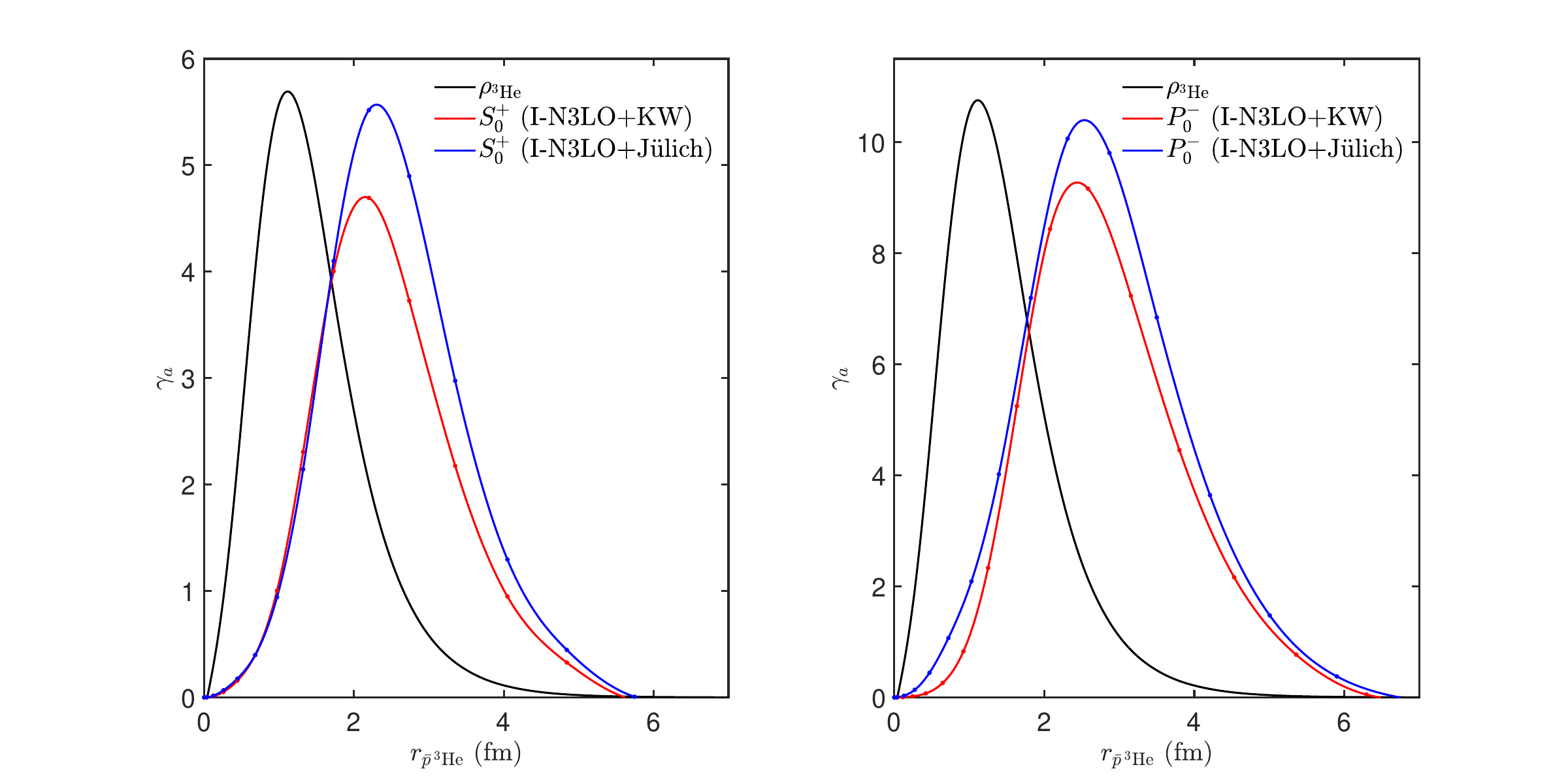}
\caption{Annihilation densities of $0^+$ (left) and $0^-$ (right) $\bar{p} \, \rm{^3He}$ states computed with the KW (red) and J\"ulich (blue) potentials, used in conjunction with the I-N3LO interaction. The curves are compared with the appropriately scaled density of $\rm{^3He}$ (black).}
\label{fig:g3He}
\end{figure}

The annihilation densities for the $0^+$ and $0^-$ states computed with different $N \bar{N}$ optical potentials are shown in Figs. \ref{fig:g3H} and \ref{fig:g3He} for $\rm{^3H}$ and $\rm{^3He}$, respectively. Each curve is plotted as a function of the antiproton–nucleus center-of-mass distance and compared with the appropriately scaled radial nuclear density (black curves). For both nuclei and both states, the overall shape of the annihilation density is largely insensitive to the choice of optical potential: only the normalization
-- proportional to the total width -- varies in accordance with the differences observed in the level shifts.

The results for $\rm{^3H}$ and $\rm{^3He}$ are qualitatively similar, as expected given that the relevant Bohr radii exceed the range of the strong interaction by nearly one order of magnitude. Most importantly, in all cases, the maximum of the annihilation density is systematically shifted outward by roughly $1 \, \rm{fm}$ relative to the peak of the nuclear density. This clearly demonstrates that the antiproton annihilation occurs predominantly in the nuclear periphery, thereby supporting the central premise of the PUMA project. A comparable behaviour is also found for other $S$ and $P$ states.

\section{Conclusion}

As a contribution to the theoretical foundation of the PUMA experiment at CERN, we have performed the first \textit{ab initio} calculations for four-body antiproton-nucleus systems. In particular, we computed the scattering lengths and volumes for the $\bar{p}+\rm{^3H}$ and $\bar{p}+\rm{^3He}$ systems and determined the level shifts of the corresponding hydrogenic states  by solving the Faddeev-Yakubovsky equations in configuration space. The sensitivity of these observables to the underlying strong interactions was systematically investigated by employing several realistic $NN$ and $N \bar{N}$ interaction models. 

In all cases, the influence of the $NN$ interaction was found to be negligible. By contrast, while different 
$N \bar{N}$ models yield mutually consistent predictions for $S$-wave states, the energies of the hyperfine-split $P$ states exhibit pronounced model dependence. The substantial discrepancies among predictions based on realistic $N \bar{N}$ interactions, together with the observed deviation from the experimentally measured  $2p$ shift, highlight the present limitations in our understanding of the low-energy $N \bar{N}$ interaction and underscore the need for additional experimental constraints in this sector. 

Finally, by computing annihilation densities for low-lying hydrogenic states, we have demonstrated-—largely independent of the underlying interaction model -- that \textit{ab initio} calculations robustly support the peripheral nature of antiproton absorption. This behaviour constitutes a central assumption of the PUMA strategy for probing the tails of proton and neutron density distributions. Our results therefore provide firm theoretical validation for this approach and reinforce the potential of antiprotonic atoms as precision tools for studying nuclear surface structure.

\section{Acknowledgement}
This work has received funding from the F.R.S.-FNRS under Grant No. 4.45.10.08. Computational resources have been provided by the Consortium des \'Equipements de Calcul Intensif (C\'ECI), funded by the Fonds de la Recherche Scientifique de Belgique (F.R.S.-FNRS) under Grant No. 2.5020.11 and by the Walloon Region. In preparing this manuscript we have benefited from the grant of French CNRS/IN2P3 for a theory project “PUMA”. We also have been granted access to the HPC resources of TGCC/IDRIS under the allocation AD010506006 made by GENCI (Grand Equipement National de Calcul Intensif).


\providecommand{\noopsort}[1]{}\providecommand{\singleletter}[1]{#1}%

\end{document}